\documentclass[aps,prl,twocolumn,superscriptaddress,twoside,floatfix,amsmath,showpacs]{revtex4}

\usepackage{graphicx}

\newcommand{\cerusi}{CeRu$_2$Si$_2$}
\newcommand{\threetwoseven}{Sr$_3$Ru$_2$O$_7$}

\begin{document}

\title{Continuous Evolution of the Fermi Surface of \cerusi\\ across
  the Metamagnetic Transition}

\author{R. Daou}
\affiliation{Cavendish Laboratory, University of
  Cambridge, Madingley Road, Cambridge CB3 0HE, UK.}

\author{C. Bergemann}
\affiliation{Cavendish Laboratory, University of
  Cambridge, Madingley Road, Cambridge CB3 0HE, UK.}

\author{S. R. Julian}
\affiliation{Cavendish Laboratory, University of
  Cambridge, Madingley Road, Cambridge CB3 0HE, UK.}
\affiliation{Department of Physics, University of Toronto, Toronto M5S
  1A7, Ontario, Canada}

\date{Submitted to Physical Review Letters on 31 May 2005}

\begin{abstract}
  We present new, high resolution Hall effect and magnetoresistance
  measurements across the metamagnetic transition in the heavy fermion
  compound \cerusi. The results force us to rethink the notion that
  the transition is accompanied by an abrupt f-electron localisation.
  Instead, we explain our data assuming a {\em continuous\/} change of
  the Fermi surface. We also point out ambiguities in the
  interpretation of dHvA data and give a possible solution to the
  problem of the ``missing mass''.
\end{abstract}


\maketitle

The nature and mechanism behind metamagnetic transitions (MMTs) in
itinerant electron systems has recently received a surge of interest,
partly due to the observation of quantum critical phenomena in
metamagnetic materials such as \threetwoseven\ 
\cite{perry01,grigera01}. In this context, the intriguing MMT in the
heavy fermion metamagnet \cerusi\ is now being revisited after two
decades of intensive research \cite{flouquet02}. \cerusi\ possesses a
massively enhanced specific heat coefficient $\gamma \simeq
350$\,mJ/mole\,K$^2$ \cite{besnus85thompson85} and shows no
superconducting or magnetic order in zero field---but an applied field
of around $8\,{\rm T}/\cos\theta$ induces a MMT into a strongly
spin-polarised state \cite{haen87} (the dependence on the angle
$\theta$ between the field and the $c$-axis reflects the Ising
character of the f-moments that underlie the Abrikosov-Suhl
resonance).

The metamagnetic effect in \cerusi\ is accompanied by the general
tendency of an applied magnetic field to split and suppress the Kondo
resonance and regain the localised f-electron \cite{costi00}.  In
fact, the present view appears to be that the f-electron abruptly
localises {\em at\/} the MMT, a notion which is largely based on
interpretation of de Haas-van Alphen (dHvA) data \cite{aoki93}.  This
is despite a number of obvious problems with this viewpoint: (a) The
MMT appears to be a crossover rather than a real transition
\cite{flouquet02}, while Luttinger's theorem applied to f-localisation
necessitates a discrete change in Fermi surface volume.  (b) On the
high-field side of the MMT, both the magnetisation ($0.7\,\mu_B$ per
Ce \cite{haen87}) and the specific heat (500\,mJ/mole\,K$^2$
\cite{vandermeulen91}) depart drastically from those of the local
moment \cite{king91} analogue CeRu$_2$Ge$_2$ ($2.15\,\mu_B$ and
20\,mJ/mole\,K$^2$ \cite{bohm88}).  The differences lessen somewhat
away from the MMT, but large discrepancies persist up to the highest
fields in which \cerusi\ has been studied \cite{vandermeulen91}. (c)
Analysis of the dHvA data within the 4f localisation scenario can in
fact only account for 20\,\% of the observed specific heat
\cite{tautz95}, i.e.\ there is ``missing mass''. (d) Finally, the
whole concept of a ``large'' Fermi surface that does include the
f-electrons, vs.\ a ``small'' one that does not, becomes blurred for
spin-polarised systems.  Mathematically, the only criterion to
distinguish ``large'' and ``small'' Fermi surfaces is whether the
Fermi volume contains an odd or even number of electrons.  If
$\uparrow$- and $\downarrow$-electrons are different, one has to count
each spin flavour separately, and the Fermi volume can change by one
electron at a time without invoking f-localisation.  [Here the
``change'' in Fermi volume is largely due to accountancy changes:
e.g.\ when hitherto disregarded bands below the Fermi level start to
form small hole pockets.]

If the ``large/small'' localisation scenario is inappropriate for
spin-polarised \cerusi\ at its MMT, the question remains as to what
actually happens to the electronic structure there.  The issue has
been dormant for a while, but there are a number of recent, 
related results ascribing anomalies in the transport properties or in
the magnetic response in other materials to changes or complete
breakdown of their Fermi surface topology
\cite{paschen04,lee04,kikugawa04}. 

Here we present magnetoresistance and Hall data on a purer \cerusi\ 
sample, down to lower temperatures, over a wider field range than had
previously been possible. Analysis of our data within an orbital
magnetoconductivity model suggests that \cerusi\ {\em continuously\/}
loses one spin-split Fermi sheet at the MMT, while the overall Fermi
volume remains constant. This resolves the difficulties of the
f-localisation scenario.  We also reassess dHvA data and point out
interpretational ambiguities, to solve the riddle of the missing mass.

\paragraph{Experiment}

The \cerusi\ samples were grown by F.~S.~Tautz using the Czochralski
technique \cite{tautz95}. One rectangular platelet of dimensions
0.65$\times$0.5$\times$0.1\,mm, the short dimension along the
$c$-axis, was soldered to at the four corners, with contact
resistances estimated to be less than 0.1\,m$\Omega$ at low $T$.
We made AC four-terminal resistance measurements with sensing
currents 1.6\,mA (for $T \ge 100$\,mK) and 0.16\,mA ($T <
100$\,mK), at 77\,Hz, using low-temperature transformers, on a
dilution refrigerator with a base $T$ of 6.5\,mK, in magnetic
fields up to 16\,T.

The in-plane transverse magnetoresistivity ($\jmath \| ab,\ B \| c$) was
determined using the Montgomery geometry factor for rectangular sample
dimensions, refined for our slightly irregularly-shaped sample through
explicit numerical solution of Laplace's equation over the sample
area. Similarly, the Hall resistivity was measured in the ``diagonal''
four-terminal geometry, here we used field-inversion and
antisymmetrisation to further reduce the error arising from geometric
uncertainties.

\paragraph{Results}

\begin{figure}[tbp]
\includegraphics[width=\columnwidth]{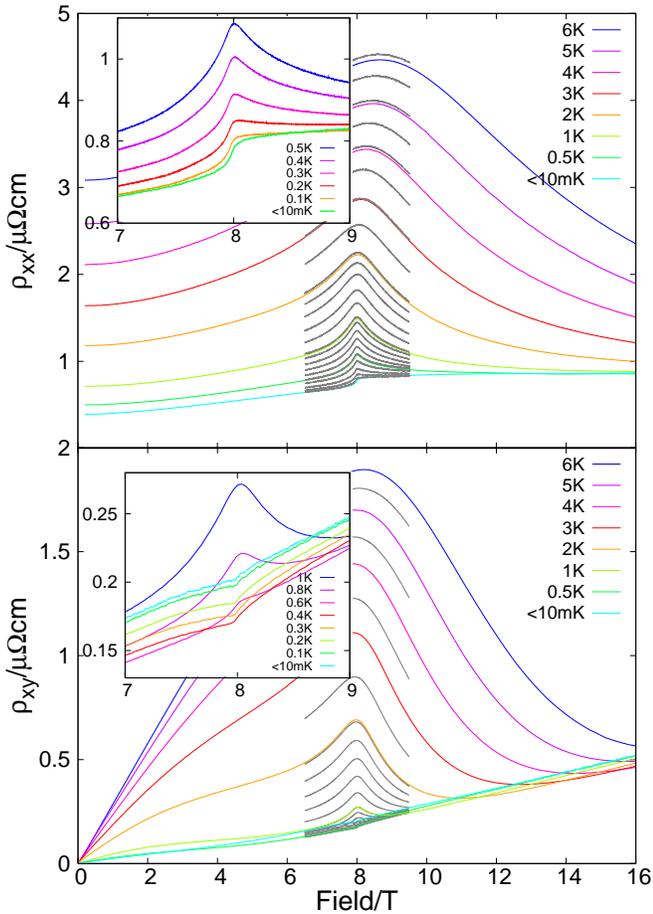}
\caption{In-plane ($\jmath \| ab,\ B \| c$) transverse
  magnetoresistivity ($\rho_{xx}$, upper panel) and Hall effect
  ($\rho_{xy}$, lower panel) for \cerusi. The base $T$ is
  around 6.5\,mK.}
\label{traces}
\end{figure}

Fig.~\ref{traces} shows the in-plane transverse magnetoresistivity
$\rho_{xx}$ as a function of temperature and field in the upper panel,
and the Hall resistivity $\rho_{xy}$ in the lower panel. The in-plane
residual resistivity is around 0.4\,$\mu\Omega$\,cm.  Our results are
broadly similar to previous magnetoresistance
\cite{kambe95,weickert05} and Hall effect \cite{kambe96} studies, but
we have extended them to a much wider field range and down to lower
temperatures, revealing new features that necessitate qualitative
changes in interpretation.

In particular, the peak in $\rho_{xx}$ at the MMT, associated with
magnetic scattering, initially sharpens up when $T$ is lowered, but at
the lowest temperatures the peak turns into a kink. A similar peak in
$\rho_{xy}$ disappears completely and turns into a {\em downward} kink
at the lowest temperatures---a feature entirely missed in the previous
studies \cite{kambe96}, demonstrating a much improved signal-to-noise
ratio. At low-$T$, $\rho_{xy}$ in our measurements increases
monotonically with field and lacks the unphysical plateau regions that
were previously observed \cite{kambe96}.

\paragraph{Resistivity Power Laws} 

A representation of the effect of magnetic fluctuation scattering on
the resistivity near a magnetic critical point can be obtained through
power law analysis. The logarithmic derivative $x =
\partial\ln(\rho-\rho_0) / \partial\ln T$ reveals whether the system
is in a Fermi liquid region with dominant electron-electron scattering
($x=2$) or whether stronger fluctuation effects dominate. Such power
law analysis has directly revealed the phase diagram in
\threetwoseven\ \cite{perry01}, YbRh$_2$Si$_2$ \cite{custers03}, and
La$_y$Sr$_{2-y}$RuO$_4$ \cite{kikugawa04}.

\begin{figure}[tbp]
\includegraphics[width=\columnwidth]{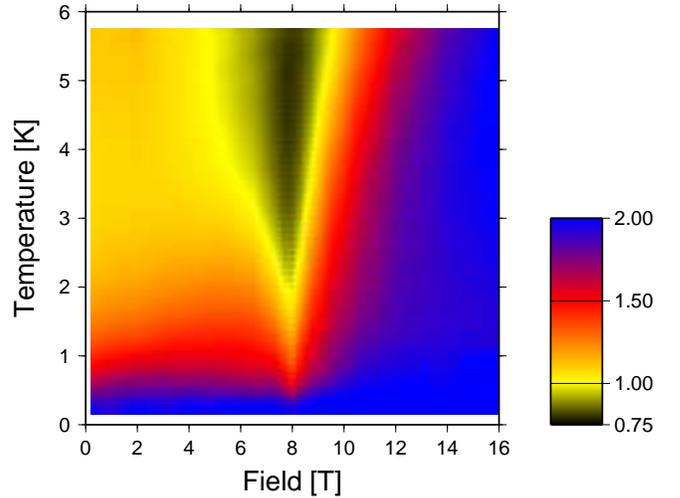}
\caption{Resistivity power law exponent $x$ versus field and
  temperature, interpolated from temperature sweeps at 0.2, 2, 4, 5.5,
  6.5, 7.3, 7.7, 8, 8.3, 8.7, 9.5, 10.5, 12, 14, and 16\,T. The Fermi
  liquid ($x=2$) region is strongly suppressed in the heavy fermion
  state below the MMT.}
\label{expo}
\end{figure}

We have have performed additional temperature sweeps at fixed magnetic
field to facilitate this kind of analysis.  Fig.~\ref{expo} reveals
how the Fermi liquid state is confined to a low $T$ region below the
MMT and then rapidly but continuously opens out above 8\,T\@. This is
a striking qualitative difference to \threetwoseven\ \cite{perry01}
where, broadly speaking, the situation is the opposite, and the
high-field state is the more renormalised one. At the MMT in \cerusi,
$\rho_{xx}(T)$ actually behaves sub-linearly over a wide temperature
range, with an effective power law exponent around $x = 0.75$.  The
$A$-coefficient in the resistivity is roughly reciprocal to the width
of the blue (Fermi liquid, $x=2$) region in Fig.~\ref{expo} and tracks
both $\gamma^2$ \cite{vandermeulen91} and the $1/T_1T$ in $^{99}$Ru
NMR \cite{ishida98} extremely well.

\paragraph{Orbital Analysis} 

The most important feature in our data is the absence of a
discontinuity in either the magneto- or the Hall resistivity as the
MMT is crossed. If the MMT represented an abrupt 4f localisation
transition, a discontinuity would almost certainly be expected since,
in the simplest view, the Hall number tracks the carrier density,
while the resistivity tracks the Fermi surface area. Also,
localization would turn \cerusi\ into an uncompensated metal, implying
huge changes in the magnetoresistance which we do not observe. [In
more sophisticated orbital magnetoresistance calculations, the exact
shape of the expected discontinuity depends on the Fermi surface
geometry, but a sudden change in one still implies an abrupt
discontinuity in the other.]

It is possible, however, to interpret the low-$T$ magnetoresistance
and Hall curves in Fig.~\ref{traces} in terms of a less drastic Fermi
surface topology transition. For this, we first need to separate the
orbital contributions to the Hall effect from the anomalous ones which
arise from skew electron-electron and electron-impurity scattering.
For the magnetoresistance, extracting the orbital part is simple: we
just need to view the lowest-$T$ data where electron-electron
scattering has been cut out. It can be more elaborate to extract the
orbital contribution of the Hall effect: the usual analysis
\cite{paschen04} assumes that the $T$-dependent low-field Hall
coefficient depends linearly on $\rho\chi$ \cite{fert87}, so that an
extrapolation to $\rho\chi = 0$ gives the orbital part. In our data,
this linear relationship appears to be invalid, so we have to
resort to a more qualitative discussion. The total low-$T$ Hall
coefficient is around $2\times 10^{-9}\,{\rm m}^3/{\rm C}$, roughly 
the right magnitude expected for the orbital effect; in comparison,
the anomalous contribution in other heavy fermion systems
near magnetic instabilities such as YbRh$_2$Si$_2$ \cite{paschen04} is
less than $0.1\times 10^{-9}\,{\rm m}^3/{\rm C}$ even for residual
resistivities 2--3 times higher than in our sample. We can therefore
conclude with some confidence that the bulk of our lowest-$T$ Hall
signal is orbital, not anomalous, in origin.

We now have to explain the kinks in both $\rho_{xx}$ and $\rho_{xy}$.
Such kinks can be the signature of weaker electronic topological
(sometimes called Lifshitz, or 2$\frac{1}{2}$) transitions that are
associated with the Fermi level moving through a van Hove singularity
in the density of states of a 3D metal \cite{blanter94}.  From a Fermi
surface point of view, these van Hove singularities correspond to the
band edges of the strongly renormalised f-band, i.e.\ to the point
where the heavy $\psi$-surface disappears. In other words, the Zeeman
spin-splitting of the heavy $\psi$-surface leads to progressive
shrinking of the majority spin $\uparrow$-sheet (recall that $\psi$ is
a hole sheet). If we look at the relevant energy scales, the
$\psi$-surface has a Fermi temperature of order the Kondo temperature,
$T_K \simeq 20$\,K.  With a Wilson ratio of $\sim 2$ typical for Kondo
systems, the MMT field of 8\,T indeed appears sufficient to drive the
whole of the $\uparrow$-band below the Fermi energy so that the
$\uparrow$-volume takes up the whole BZ and its surface shrinks to a
point. Above the MMT, the heavy $\uparrow$-electron is then no longer
itinerant.

\begin{figure}[tbp]
\includegraphics[width=0.8\columnwidth]{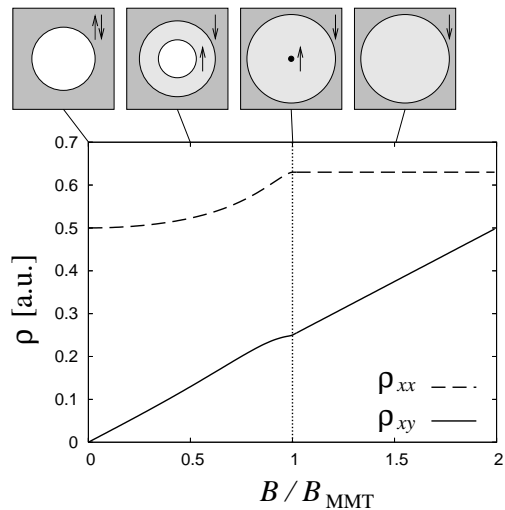}
\caption{Magneto- and Hall resistivity of a model system consisting of
  a spin-split, hole-like spherical Fermi surface where one of the
  spin sheets shrinks to a point at $B_{\rm MMT}$. The pictures show
  the field evolution of the Fermi surface.}
\label{model}
\end{figure}

While \cerusi\ is of course a multi-band metal, the essence of the
above ideas is captured by a spherical, spin-split Fermi surface in a
simple parabolic band.  To quantify the spin-splitting,
$k_{F\downarrow}^2 - k_{F\uparrow}^2$---essentially the Zeeman
splitting---is taken proportional to $B$ while the total Fermi volume
is held constant. The conductivity contributions $\sigma_{xx} \propto
k_F^2/(1+\cot^2 \phi_H)$ and $\sigma_{xy} \propto
k_F^2\cot\phi_H/(1+\cot^2 \phi_H)$ of both sheets can be added, and
the resulting resistivities are shown in Fig.~\ref{model}. $\phi_H$ is
the Hall angle: $\cot \phi_H = eBl_{\rm free} /\hbar k_F$, and the
mean free path $l_{\rm free}$ forms the only free parameter of the
model.

The comparison with the actual data in Fig.~\ref{traces} shows very
good qualitative agreement: the main features in the low-$T$ data are
remarkably well reproduced, such as the direction and relative size of
the kinks at $B_{\rm MMT}$, and the formation of the high-field
plateau in the magnetoresistivity. A more quantitative comparison is
made difficult by the presence of the other bands, and would require
much more sophisticated calculations.

One peculiarity of our data is that no further Fermi surface evolution
seems to take place beyond the MMT.  A previous scenario
\cite{edwards97} suggested that the $\uparrow$-level should go {\em
  through\/} the f-level hybridization gap and create a small but
expanding electron sheet on the high-field side of the MMT. Additional
calculations that we have done show that this new electron sheet
should have a dramatic influence on the Hall effect, a feature that is
entirely absent in the data. Our experiment therefore forces us to
conclude that the $\uparrow$-level remains within the hybridization
gap for fields up to at least 16\,T\@.

The presence of a hybridization gap appears to conflict with the
observation of enhanced specific heat \cite{vandermeulen91} and
differential susceptibility at the MMT which indicates a peak, not a
gap, in the density of states. However, it has been suggested that
fluctuation contributions can further enhance the specific heat near
Fermi surface topology transitions \cite{blanter94}, producing an inverse
square root divergence at the MMT. The observed specific heat data
\cite{vandermeulen91} does in fact resemble such a fluctuation peak
superposed on a band edge. Also, the lattice softening at the
transition \cite{flouquet02} already points towards significant
effects of coupled volume and magnetisation fluctuations on the
thermodynamic properties.

\paragraph{DHvA Fermi Surface Data}

At first glance, existing \cite{takashita96} and our own dHvA data
seem at odds with the continuous Fermi surface evolution discussed
above, since a frequency of 28\,kT is observed which matches a large
hole orbit ($\omega$) in band calculations \cite{yamagami92} that
assume complete f-localisation. This constitutes the main evidence for
the existing, abrupt ``large/small'' transition scenario
\cite{aoki93}.

However, $\omega$ is in fact one of {\em two\/} orbits with roughly
the same dHvA frequency for on-axis fields, the other being a hole
orbit ($\mu$) in the extended sheet of the f-itinerant band structure
which is applicable at the very least to the low-field state of \cerusi.
This accidental degeneracy of dHvA frequencies has been somewhat
overlooked, even though the $\mu$-orbit is actually seen,
unambiguously and {\em below\/} the MMT, when \cerusi\ is pressurised
\cite{aoki01}.  Above the MMT, the 28\,kT branch disappears for
$\theta > 10^\circ$ \cite{takashita96}, which is indeed expected for
$\mu$ from the quantitative band structure \cite{yamagami92}---in
contrast, $\omega$ should be visible much more clearly away from the
$c$-axis, as it is e.g.\ in CeRu$_2$Ge$_2$ \cite{king91}, due to its
smaller off-axis cross-section and more favourable curvature.  Also,
if the 28\,kT frequency is reassigned to the $\mu$-orbit, its moderate
mass enhancement is exactly what is expected from comparison with
other orbits like $\kappa/\delta$ on the same sheet ($m^\star/m \simeq
10$).

\begin{figure}[tbp]
\includegraphics[width=0.8\columnwidth]{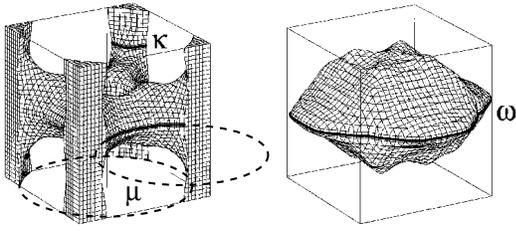}
\caption{The $\mu$-orbit in the f-band calculation (left, centered on
  $\Gamma$) and $\omega$-orbit in the f-core calculation (right,
  centered on Z) in \cerusi\ \cite{yamagami92} have very similar on-axis dHvA
  frequencies, around 28\,kT. Figure adapted from
  Ref.~\onlinecite{aoki01}.}
\label{orbits}
\end{figure}

Therefore, while the MMT certainly induces slight changes in the
observed (back-projected) dHvA frequencies on the minor sheets
\cite{takashita96,aoki01}, the existing dHvA data can be brought in
line with the continuous Fermi surface transition scenario. In this,
the heavy $\downarrow$-orbit remains unobserved---but this is not too
worrisome since (as $\psi$) it is quite elusive even below the MMT
unless the field is exactly in the plane. This orbit carries most of
the density of states and accounts for the ``missing mass'' that
presented such a puzzle in the f-localisation scenario.

The fluctuation scenario invoked in the previous section helps to
explain some of the remaining puzzles in the dHvA data. At the MMT,
the dHvA signal is strongly damped even at the lowest temperatures,
but this damping is not accompanied by a peak in the low-$T$
resistivity (see Fig.~\ref{traces}). This points to the importance of
low-$k$ fluctuations which destroy quantum coherence but do not affect
transport properties. The anomalous damping, and enhanced temperature
sensitivity of the dHvA effect, is seen on {\em all\/} sheets of the
Fermi surface, not just on the heavy orbit, suggesting that the
specific heat divergence is not purely a quasi-1D band edge effect,
but again that fluctuations play the main role.

In conclusion, our new interpretation of the MMT as a continuous,
Fermi surface volume conserving, topology driven transition deals
with all the issues (a)--(d) mentioned in the introduction. Our data
forcefully agree with the notion that the MMT is actually a crossover
and that \cerusi\ remains a Fermi liquid at all fields, so a
formulation in terms of continuous Fermi surface transitions appears
as the natural and indeed the only viable viewpoint. Our analysis
suggests that the $\uparrow$-surface shrinks to a point at the
MMT---essentially recasting previous ideas \cite{edwards97} in a
continuous fashion---and it precludes any further Fermi surface
topology changes up to 16\,T.

\acknowledgments

We benefited from discussions with J. Flouquet, P. Gegenwart, K.
Ishida, D. E. Khmelnitskii, G. G. Lonzarich, A. P. Mackenzie, and D.
van der Marel. This work was funded by the U.K.\ EPSRC. C.B.\ 
acknowledges the support of the Royal Society.


\begin{thebibliography}{99}

\bibitem{perry01} R. S. Perry {\it et al.}, Phys.\ Rev.\ Lett.\ {\bf
    86}, 2661 (2001).

\bibitem{grigera01} S. A. Grigera {\it et al.}, Science {\bf
    294}, 329 (2001).

\bibitem{flouquet02} J. Flouquet {\it et al.}, Physica B {\bf 319},
  251 (2002).
  
\bibitem{besnus85thompson85} M. J. Besnus {\it et al.}, Solid State
  Commun.\ {\bf 55}, 779 (1985); J. D. Thompson {\it et al.}, {\it
    ibid.} {\bf 56}, 169 (1985).

\bibitem{haen87} P. Haen {\it et al.}, J.\ Low Temp.\ Phys.\ {\bf 67},
  391 (1987).

\bibitem{costi00} T. A. Costi, Phys.\ Rev.\ Lett.\ {\bf 85}, 1504 (2000).

\bibitem{aoki93} H. Aoki {\it et al.}, Phys.\ Rev.\ Lett. {\bf 71},
  2110 (1993). 

\bibitem{vandermeulen91} H. P. van der Meulen {\it et al.}, Phys.\
  Rev.\ B {\bf 44}, 814 (1991).

\bibitem{king91} C. A. King and G. G. Lonzarich, Physica B {\bf 171},
  161 (1991).

\bibitem{bohm88} A. Bohm {\it et al.}, J. Magn.\ Magn.\ Mat.\ {\bf
    76}, 150 (1988).

\bibitem{tautz95} F. S. Tautz {\it et al.}, Physica B {\bf 206}, 29
  (1995).

\bibitem{paschen04} S. Paschen {\it et al.}, Nature {\bf 432}, 881
  (2004). 

\bibitem{lee04} M. Lee  {\it et al.}, Phys.\ Rev.\ Lett. {\bf 92},
  187201 (2004).

\bibitem{kikugawa04} N. Kikugawa {\it et al.}, Phys. Rev. B {\bf 70},
  134520 (2004).

\bibitem{kambe95} S. Kambe {\it et al.}, Solid State Commun.\ {\bf
    96}, 175 (1995).

\bibitem{weickert05} F. Weickert {\it et al.}, Physica B (in press).

\bibitem{kambe96} S. Kambe {\it et al.}, J.\ Low Temp.\ Phys.\ {\bf
    102}, 477 (1996).

\bibitem{custers03} J. Custers {\it et al.}, Nature {\bf 424}, 504 (2003)

\bibitem{ishida98} K. Ishida {\it et al.}, Phys. Rev. B {\bf 57},
  11\,054 (1998). 

\bibitem{fert87} A. Fert and P. M. Levy, Phys. Rev. B {\bf 36}, 1907
  (1987). 

\bibitem{edwards97} D. M. Edwards and A. C. M. Green, Z.\ Phys.\ B
  {\bf 103}, 243 (1997).
  
\bibitem{blanter94} For a review, see Y. M. Blanter {\it et al.},
  Phys.\ Rep.\ {\bf 245}, 159 (1994).

\bibitem{takashita96} M. Takashita {\it et al.}, J.\ Phys.\ Soc.\
  Jpn.\ {\bf 65}, 515 (1996).
  
\bibitem{yamagami92} H. Yamagami and A. Hasegawa, J.\ Phys.\ Soc.\ 
  Jpn.\ {\bf 61}, 2388 (1992); {\it ibid.} {\bf 62}, 592 (1993).

\bibitem{aoki01} H. Aoki {\it et al.}, J.\ Phys.\ Soc.\ Jpn.\ {\bf
    70}, 774 (2001).
  
\end{thebibliography}
\end{document}